\documentclass[conference]{IEEEtran}
\IEEEoverridecommandlockouts

\usepackage{cite}
\usepackage{amsmath,amssymb,amsfonts}
\usepackage{algorithmic}
\usepackage{graphicx}
\usepackage{textcomp}
\usepackage{xcolor}
\usepackage{multirow}
\def\BibTeX{{\rm B\kern-.05em{\sc i\kern-.025em b}\kern-.08em
    T\kern-.1667em\lower.7ex\hbox{E}\kern-.125emX}}
\begin{document}

\title{Improving Speech Enhancement by Cross- and Sub-band Processing with State Space Model \\

\thanks{$^*$Corresponding author.}
}

\author{\IEEEauthorblockN{\textit{Jizhen Li$^{1}$, Weiping Tu$^{1,2,*}$, Yuhong Yang$^{1,2}$, Xinmeng Xu$^{1}$, Yiqun Zhang$^{1}$, Yanzhen Ren$^{3}$}}
\IEEEauthorblockA{$^1$NERCMS, School of Computer Science, Hubei Luojia Laboratory, Wuhan University, China\\
$^2$Hubei Key Laboratory of Multimedia and Network Communication Engineering, Wuhan University, China\\
$^3$School of Cyber Science and Engineering, Wuhan University, China\\
\texttt{lijizhen@whu.edu.cn, tuweiping@whu.edu.cn}}
}


\maketitle

\begin{abstract}
Recently, the state space model (SSM) represented by Mamba has shown remarkable performance in long-term sequence modeling tasks, including speech enhancement. However, due to substantial differences in sub-band features, applying the same SSM to all sub-bands limits its inference capability. Additionally, when processing each time frame of the time-frequency representation, the SSM may forget certain high-frequency information of low energy, making the restoration of structure in the high-frequency bands challenging. For this reason, we propose Cross- and Sub-band Mamba (CSMamba). To assist the SSM in handling different sub-band features flexibly, we propose a band split block that splits the full-band into four sub-bands with different widths based on their information similarity. We then allocate independent weights to each sub-band, thereby reducing the inference burden on the SSM. Furthermore, to mitigate the forgetting of low-energy information in the high-frequency bands by the SSM, we introduce a spectrum restoration block that enhances the representation of the cross-band features from multiple perspectives. Experimental results on the DNS Challenge 2021 dataset demonstrate that CSMamba outperforms several state-of-the-art (SOTA) speech enhancement methods in three objective evaluation metrics with fewer parameters.
\end{abstract}


\section{Introduction}

Speech signals are often subjected to various environmental noises, which significantly degrade the quality and intelligibility of the speech signal. Speech enhancement aims to improve the clarity and perceptibility of the target speech in the presence of environmental noise interference. It plays a crucial role in various applications, including hearing aids~\cite{edwards2007future}, automatic speech recognition (ASR)~\cite{malik2021automatic}, audio broadcasting~\cite{faller2002technical} and speech processing~\cite{pandey2018adversarial}.

Comprehensive methods combining Transformer with CNN and RNN have been proposed in the speech enhancement task~\cite{abdulatif2024cmgan, li24w_interspeech, xu2023case, 10214650}, to address the limitations that CNN-based methods have limited receptive fields and RNN-based approaches exhibit limited memory capacity. Transformer-based methods have demonstrated remarkable performance in modeling long-range dependencies in speech signals~\cite{xu2024adaptive, chen20l_interspeech, 10176306}, but they often suffer from quadratic computational complexity and high hardware requirements~\cite{xu23f_interspeech, chen2024mim, Xu_Tu_Yang_2023, jiang2024dual}. Although linear Transformers have been introduced to mitigate this problem~\cite{han2023flatten}, they generally exhibit performance degradation. Therefore, the main focus in speech enhancement remains on reducing complexity while maintaining high performance.

Recently, the advanced state space model Mamba based on S6~\cite{gu2023mamba} has achieved comparable or superior performance to Transformer in long-term sequence modeling tasks, while maintaining linear complexity~\cite{zhao2024sicrn, guo2024mambair, ma2024u, zhu2024vision}. SSM utilizes state transition matrix~\cite{gu2020hippo} to compute hidden states, enabling efficient compression and storage of information. Owing to the excellent inference capability of Mamba, it exhibits promising performance across many tasks. At present, the dual-path architecture, which incorporates separate processing for the cross- and sub-bands, has gained popularity in speech enhancement. However, simply applying SSM to dual-path architecture faces challenges: (1) The weight-shared state transition matrix in SSM struggles to effectively infer all sub-band features with significant differences. (2) Due to the concentration of energy in the middle and low-frequency bands in the cross-band features, a single SSM may forget the information in the high-frequency bands during the inference process, resulting in the loss of spectral structure. Therefore, a new SSM-based architecture is needed to better suit the spectral-temporal characteristics of speech for speech enhancement.

In this paper, we propose Cross- and Sub-band Mamba (CSMamba), which offers the flexibility to apply SSM to sub-bands with significant differences and preserves more spectral detail information in the cross-bands. Specifically, for sub-band features extraction, unlike the uniform sub-band splitting method used in BSRNN~\cite{luo2023music}, we employ a band split block (BSB) that splits the full-band into four sub-bands with different width and applies independent convolution to each sub-band. We also design a spectrum restoration block (SRB) that splits the features across channels to extract separate cross-band features. Additionally, we employ channel integrating block (CIB) to allocate feature weights across channels, facilitating information interaction between channels.

\begin{itemize}
\item To assist the SSM in handling sub-band features flexibly, we propose BSB to allocate separate weights to different sub-bands. BSB aims to preserve the characteristics of each sub-band and enhancing the robustness of the model.

\item To mitigate the potential loss of low-energy spectral information during inference, we propose SRB to enhance the representation of cross-band features from multiple perspectives. SRB aims to restore finer spectral structures.

\item Experimental results on the DNS-Challenge 2021 dataset~\cite{reddy2021icassp} demonstrate that our CSMamba outperforms several SOTA models based on Transformer and Mamba.
\end{itemize}

\section{PROPOSED METHOD}

\subsection{Overview}

\begin{figure}
    \centering
    \includegraphics[width=0.9\linewidth]{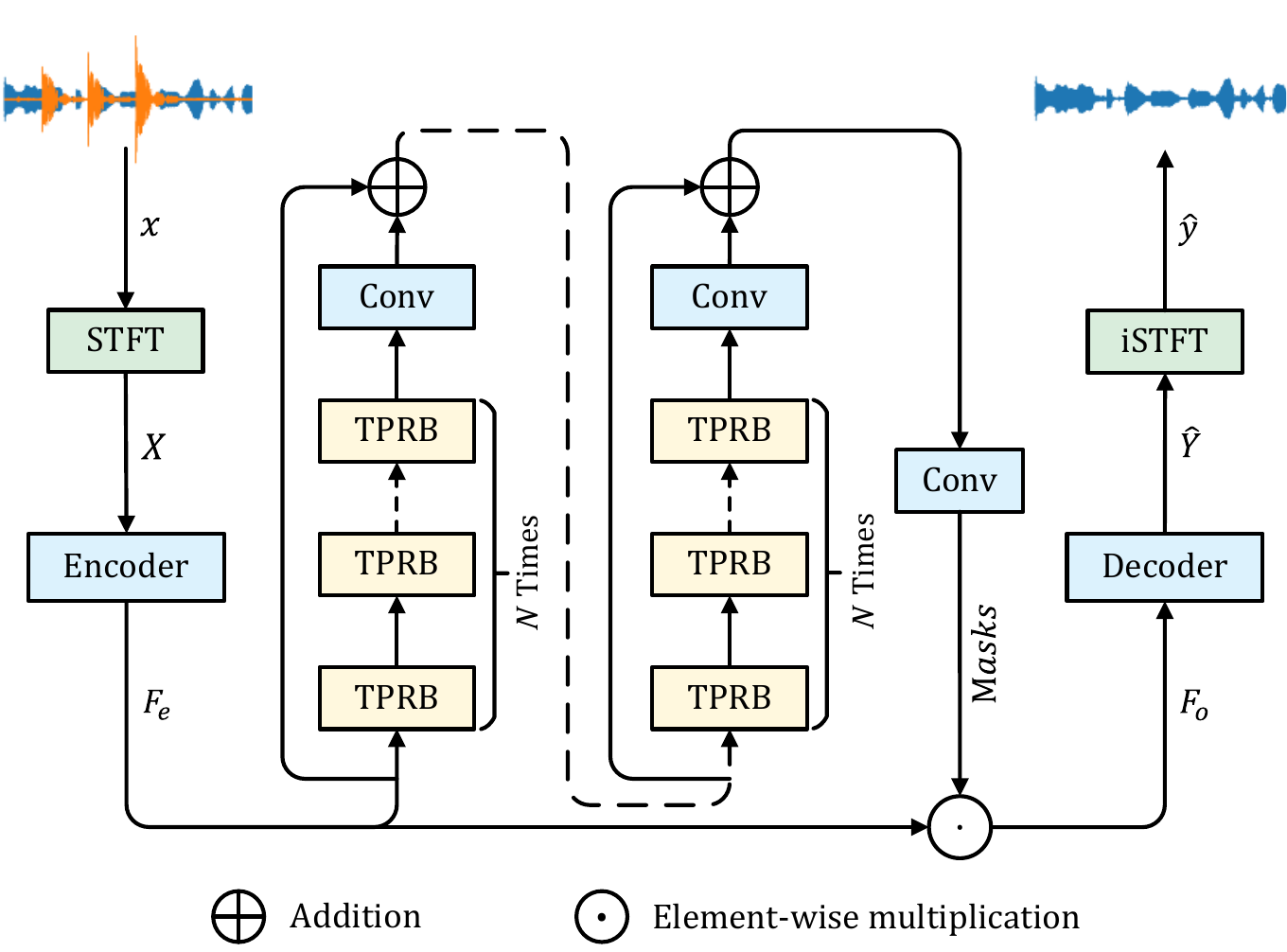}
    \caption{The overview of proposed CSMamba, which consists of $N\times L$ TPRB blocks.}
    \label{fig:CSMamba}
\end{figure}

The proposed architecture of CSMamba is illustrated in Fig.~\ref{fig:CSMamba}. The mixed signal $x$ is transformed into the complex spectrum using the short-time Fourier transform (STFT). The real and imaginary parts of the spectrum are then concatenated as the input to the model, represented as $X\in \mathbb{R}^{B\times F\times T\times 2}$, where B, F, and T denote the batch size, frequency bins, and temporal frames, respectively. The encoder preprocesses the input using convolution to obtain a high-dimensional representation $F_{e}$. The feature then passes through multiple triple-path residual blocks (TPRBs), which predict feature masks for the real and imaginary parts. The masks are then multiplied with the output $F_{e}$ from the encoder to obtain the representation of the target. Finally, the decoder generates the real and imaginary parts of the predicted target spectrogram, which are then fed into the inverse short-time Fourier transform (iSTFT) to obtain the estimated speech signal.

\subsection{Triple-Path Residual Blocks}

\begin{figure*}
    \centering
    \includegraphics[width=1.0\linewidth]{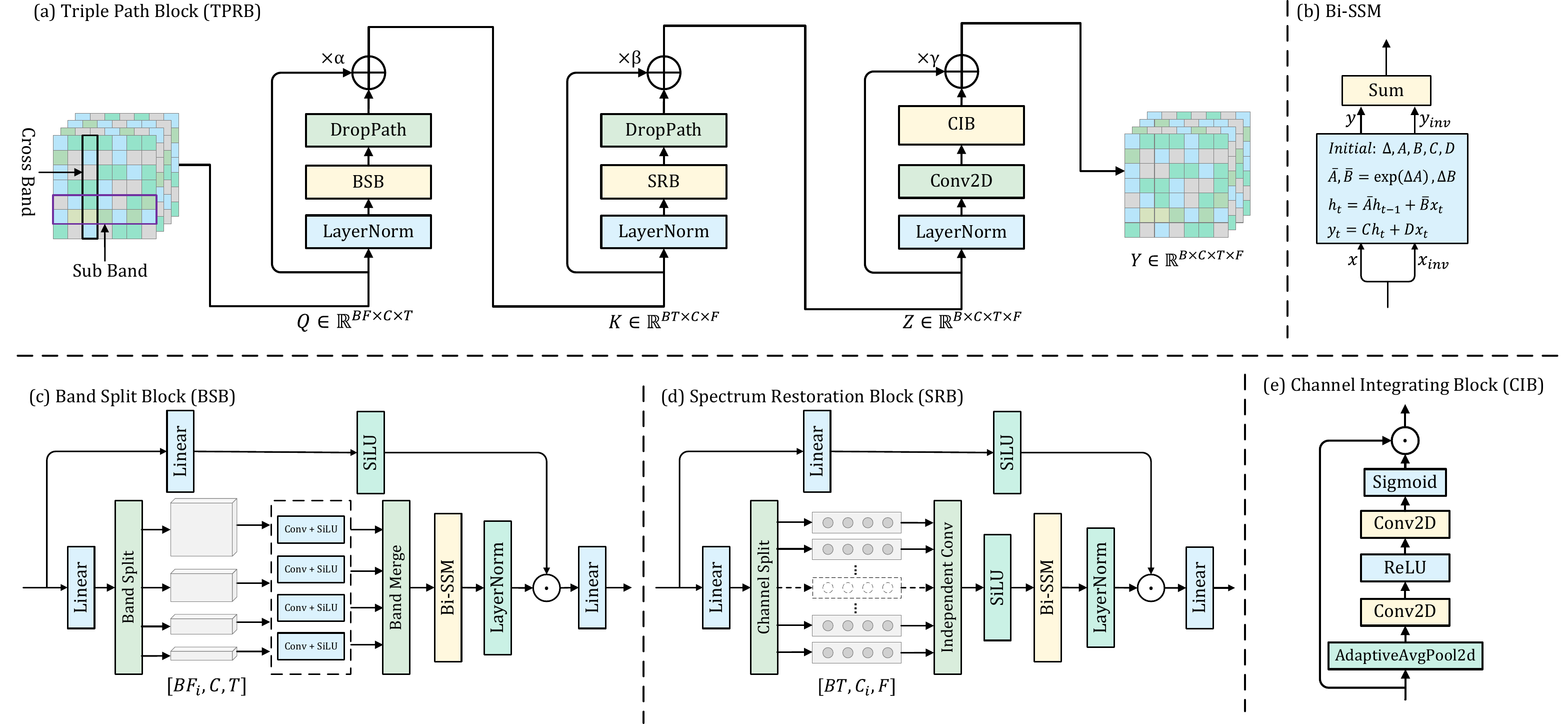}
    \caption{(a) The detail of proposed triple-path residual block. (b) The architecture of bidirectional state space model. (c) The detail of the proposed Band Split Block. (d) The detail of the proposed Spectrum Restoration Block. (e) The detail of the proposed Channel Integrating Block.}
    \label{fig:TPRB}
\end{figure*}

The core component of CSMamba consists of several cascaded TPRBs and convolutional layers. Specifically, after conducting deep feature extraction every $N$ TPRBs, a convolutional operation is applied to form a residual structure~\cite{guo2024mambair}. The residual structure is repeated $L$ times in total, resulting in a total utilization of $N\times L$ TPRBs. As shown in Fig.~\ref{fig:TPRB}, the input of the TPRB is first reshaped into $Q\in \mathbb{R}^{BF\times C\times T}$ and then processed by the Band Split Module to handle the sub-bands, where C represents the number of channels in $F_{e}$. This module extracts global features from the time dimension. Subsequently, the features are reshaped into $K\in \mathbb{R}^{BT\times C\times F}$ and fed to the Spectrum Restoration Module for processing the entire frequency bands. This module refines the restoration of spectral structural information. 

In the band split module and the spectrum restoration module, we employ bidirectional state space model (Bi-SSM) as the core processing module to fully exploit the information inherent in the speech signal. The structure of Bi-SSM is shown in Fig.~\ref{fig:TPRB}(b). The $\bar{A}$, $\bar{B}$, $C$ and $D$ are dynamic weight matrices in the network training process, where $\bar{A}$ and $\bar{B}$ are discretized from $A$ and $B$ using the zero-order hold technique by introducing a time step $\Delta$~\cite{gu2023mamba}. We adopt a lightweight bidirectional mechanism to process only the core SSM, and the formula is as follows:
\begin{equation}
    \begin{matrix}x_{\to}=x_{in}, x_{\gets}=Flip(x_{in}), 
    \\x_{out}=SSM(x_{\to}) + Flip(SSM(x_{\gets})),
    \end{matrix}
\end{equation}
where $x_{\to}$ and $x_{\gets}$ represent forward and reverse information, respectively. By introducing a bidirectional mechanism, we compensates for the parts affected by the sequence modeling order, thereby enhancing its performance. Finally, the features are transformed back into the 2D representation $Z\in \mathbb{R}^{B\times C\times T\times F} $, utilizing Channel Attention to facilitate cross-channel interaction of T-F information and reassign the weights of each channel. 

\subsubsection{Band Split Block}

The distribution of information in speech varies across sub-bands. Specifically, the mid-low frequency range is important for speech intelligibility and it contains most of the energy. In contrast, the high-frequency components contain less energy but provide detailed speech information and texture, exhibiting significant differences compared to the mid-low bands. When applying the SSM to these sub-band features with substantial information disparities, the inference capability of the weight-sharing state transition matrix is limited. 

To address this limitation, we propose the Band Split Block (BSB), which processes sub-bands flexibly, as illustrated in Fig.~\ref{fig:TPRB}(c). Specifically, we split the full-band into four sub-bands with different widths based on the differences in human ear perception of the frequency components of speech signals and the information similarity. By employing independent convolutions on these sub-bands, this module can flexibly adjust the corresponding weights based on the information distribution characteristics of each sub-band, resulting in smoother features entering the SSM. This assists the Bi-SSM in better learning the distribution patterns of speech in the mixed features, enhancing the convergence speed and robustness of the model.

Additionally, the input of this module is passed through a linear layer and then subjected to a SiLU activation function to regulate the outflow of information~\cite{hendrycks2016gaussian, ramachandran2017searching}, forming a gating mechanism with the sub-band features processed by Bi-SSM. Finally, the information is further integrated through a linear layer to obtain the output of this module.

\subsubsection{Spectrum Restoration Block}

Speech spectrum exhibit specific structural information, such as harmonic structures. These structures provide richer sound information, making the speech sound more full and realistic. However, due to the low energy of the high-frequency components in the full-band, the SSM may overlook certain less salient spectral structural information and instead allocate more attention to the mid-low frequency components with higher energy, which may result in the loss of speech details. Therefore, we propose the Spectrum Restoration Block (SRB) to finely recover detailed structures from multiple perspectives, as shown in Fig.~\ref{fig:TPRB}(d).

Different with the band split block, the spectrum restoration block processes each time frame independently (on frequency dimension). Since an utterance exhibits similar characteristics across the entire time dimension, we share the weights of SRB across all frames. After feature extraction through a linear layer, the channel split module assign independent convolution kernels to each channel of the 1D cross-band features to obtain multi-perspective representation. Within each perspective, the cross-band information undergoes independent learning through the Bi-SSM, enabling personalized restoration of spectral structures. The SRB aiming to prevent excessive suppression of spectrum structures by a single perspective.

\subsubsection{Channel Integrating Block}

In order to facilitate effective cross-channel information interaction, we have incorporated a Channel Integrating Block (CIB) after processing the sub-band and cross-band information. By introducing a lightweight channel attention mechanism~\cite{hu2018squeeze}, the CIB captures cross-channel correlations from the results of the previous two modules and aggregates spectral structural information, as illustrated in Fig.~\ref{fig:TPRB}(e). We avoid using Mamba in the channel dimension due to the non-sequential nature of channel features. Furthermore, the channel attention mechanism employs a bottleneck structure to reassign weights to the channels of the T-F features, emphasizing the relevant channel information filtered by Mamba and disregarding irrelevant channel information, thereby effectively integrating these informative channels.

\begin{table*}[]
\centering
\setlength{\tabcolsep}{8pt}
\renewcommand{\arraystretch}{1.2}
\caption{Comparison with baseline models on DNS Challenge 2021 dataset.}
\begin{tabular}{lccccccccccc} 
\hline
\multirow{2}{*}{Model} & \multirow{2}{*}{Param.(M)} & \multirow{2}{*}{FLOPs (G)} & \multicolumn{3}{c}{-5dB}                        & \multicolumn{3}{c}{0dB}                         & \multicolumn{3}{c}{5dB}                          \\ 
\cline{4-12}
                       &                            &                            & PESQ          & STOI           & SI-SNRi        & PESQ          & STOI           & SI-SNRi        & PESQ          & STOI           & SI-SNRi         \\ 
\hline
Noisy                  & -                          & -                          & 1.29          & 61.75          & -              & 1.60          & 79.70          & -              & 1.92          & 86.45          & -               \\ 
\hline
Inter-SubNet~          & 2.29                       & 71.599                     & 2.18          & 81.00          & 12.86          & 2.65          & 88.28          & 12.22          & 3.10          & 92.42          & 10.22           \\
CMGAN~                 & 1.83                       & 62.368                     & 2.31          & 83.90          & 13.50          & 2.80          & 89.22          & 12.11          & 3.17          & 93.09          & 10.13           \\
MP-SENet~              & 2.05                       & 71.645                     & 2.39          & 84.55          & 14.83          & 2.91          & 90.18          & 13.19          & 3.25          & 93.42          & 10.74           \\
SEMamba~               & 2.25                       & 51.258                     & 2.44          & 85.33          & 15.15          & 2.97          & 90.80          & 13.51          & 3.32          & 93.91          & 11.19           \\ 
\hline
CSMamba                & \textbf{1.73}              & \textbf{35.671}            & \textbf{2.50} & \textbf{86.69} & \textbf{15.39} & \textbf{3.04} & \textbf{91.94} & \textbf{13.89} & \textbf{3.43} & \textbf{94.60} & \textbf{11.61}  \\
\hline
\end{tabular}
\end{table*}

\begin{table}[]
\centering
\label{asd}
\setlength{\tabcolsep}{10pt}
\renewcommand{\arraystretch}{1.2}
\caption{The ablation study on DNS Challenge 2021 dataset with SNR set to 0dB.}
\begin{tabular}{lccc} 
\hline
Metrics                      & PESQ          & STOI           & SI-SNRi         \\ 
\hline
CSMamba                      & \textbf{3.04} & \textbf{91.94} & \textbf{13.89}  \\ 
\hline
w/o BSB                      & 2.17          & 84.07          & 9.58            \\
\textit{- w/o band split}    & 2.87          & 90.28          & 13.02           \\
- uniform band split         & 2.89          & 91.36          & 13.65           \\ 
\hline
w/o SRB                      & 2.61          & 87.40          & 12.24           \\
\textit{- w/o channel split} & 2.77          & 89.87          & 12.39           \\ 
\hline
w/o CIB                      & 2.82          & 89.97          & 12.50           \\
\hline
\end{tabular}
\end{table}

\section{Experiments}

\subsection{Datasets}

We conducted experiments on the Deep Noise Suppression (DNS) Challenge 2021 dataset, which consists of 563 hours of clean speech clips from 2,150 speakers and 181 hours of noise clips from 150 classes. We randomly selected clean speech and noise to synthesize 35,000 mixed audio clips with SNR ranging from -5dB to 5dB. All audio clips were resampled to 16kHz, and each mixed speech clip having a duration of 4 seconds. Room impulse responses (RIRs) from the openSLR26 and openSLR28~\cite{reddy2021icassp} were randomly chosen and added to 75\% of the clean speech with a reverberation time ($T_{60}$) ranging from 0.3s to 1.3s. We selected an equal number of speech and noise clips. Both the speech and the noise clips are distinct and not duplicated, aiming to improve the model's ability to generalize to unfamiliar noise conditions.

\subsection{Architecture and Training Details}

We applied a 32ms (512 samples) Hamming window with a 16ms (256 samples) hop size for the STFT, resulting in a total of 257 frequency bins. In the TPRBs of CSMamba, we set $N=5$ and $L=4$, resulting in a total of 1.73 million parameters. Both $\alpha$, $\beta$ and $\gamma$ are trainable parameters. In the band split process, we split the full-band into four sub-bands according to the intervals $\left \{ [0, 7], [7, 65], [65, 129], [129, 257] \right \}$. We utilized a joint waveform and a loss in the T-F domain~\cite{park2022manner}, comprising an L1 loss in the time domain and a multi-resolution loss in the T-F domain. The multi-resolution loss in time-frequency domain using FFT sizes of $\left \{ 512, 1024, 2048 \right \}$ with hope size $\left \{ 50, 120, 240 \right \}$ and window length $\left \{ 240, 600, 1200 \right \}$. We chose Adam~\cite{adam} as the optimizer with an initial learning rate of 0.001. The learning rate was halved when the validation set loss did not decrease for 5 epochs.

\section{RESULTS AND DISCUSSION}

\subsection{Model Comparison}

We conducted experiments on DNS Challenge 2021 dataset with different SNR levels of -5dB, 0dB, and 5dB. We selected four baseline models for comparison, including the SOTA Mamba-based speech enhancement network SEMamba~\cite{chao2024investigation}, the latest transformer-based model MP-SENet~\cite{lu2023mp}, as well as CMGAN~\cite{abdulatif2024cmgan} and Inter-SubNet~\cite{chen2023inter}. We calculated three objective evaluation metrics on this dataset, including Wide-Band Perceptual Evaluation of Speech Quality (PESQ)~\cite{rix2001perceptual}, Short-Time Objective Intelligibility (STOI)~\cite{taal2010short} and Scale-Invariant Signal-to-Noise Ratio improvement (SI-SNRi)~\cite{le2019sdr}.

Table 1 illustrates the performance of each model on the test dataset. From the table, it can be observed that CSMamba outperforms the baseline models across all metrics with only 1.73 million parameters. Unlike Inter-SubNet, which performs poorly in low SNR environments, CSMamba exhibits promising performance across various SNRs, indicating the superior stability of the SSM-based model compared to the LSTM-based model. Compared to the SOTA transformer-based speech enhancement model MP-SENet, CSMamba has demonstrated improvements of 0.14, 1.70, and 0.71 in PESQ, STOI, and SI-SNRi, respectively. Additionally, compared to the SOTA mamba-based model SEMamba, CSMamba shows an average improvement of 0.08, 1.06, and 0.35 in terms of PESQ, STOI, and SI-SNRi, respectively. The experimental results demonstrate that our proposed architecture based on SSM exhibits better adaptability for speech enhancement tasks.

\subsection{Ablation Study}

To validate the contribution of each module in our proposed model, we conducted ablation experiments on the DNS Challenge 2021 dataset with SNR set to 0dB. We still employ PESQ, STOI, and SI-SNRi as the evaluation metrics. Table 2 illustrates the results of ablation experiments that involved without BSB, without band split, without SRB, without channel split and without CIB, which correspond to the innovative components of this study. 

Table 2 demonstrates that excluding the band split component results in a performance decrease of 0.17, 1.66, and 0.87 in the PESQ, STOI, and SI-SNRi, respectively. This indicates the significant contribution of the band split component within the entire model. Furthermore, the removal of the BSB module significantly impairs the model's performance, emphasizing its crucial role in sub-band processing. The ablation experiment conducted on the channel split operation, resulting in a decrease of 0.27, 2.07, and 1.50 in PESQ, STOI, and SI-SNRi respectively, further emphasizing the importance of our designed multi-perspective module in spectral restoration, aiding in the recovery of less prominent spectral structures in high-frequency components. Additionally, the incorporation of CIB facilitates information exchange along channels, leading to improvements of 0.22, 1.97, and 1.39 in PESQ, STOI, and SI-SNRi for the model. This indicates that there is redundancy in the features across channels after employing SSM, highlighting the significance of CIB.

\section{Conclusion}

In this paper, we present CSMamba for speech enhancement. We leverage the band split operation to assist Bi-SSM in processing sub-band features flexibly and propose SRB to restore less prominent spectral structures in the high-frequency components from multiple perspectives. Furthermore, we employ CIB to integrate information from BSB and SRB, facilitating cross-channel information interaction. Experimental results demonstrate that our model outperforms several latest baseline models with fewer parameters.

\section*{Acknowledgment}

This work is supported by the National Nature Science Foundation of China (No. 62471343, No. 62071342, No.62171326), the Special Fund of Hubei Luojia Laboratory (No. 220100019).

\bibliographystyle{IEEEtran}
\bibliography{refs}

\end{document}